# Risk-Neutral Valuation of Options under Arithmetic Brownian Motions

Qiang LIU
*School of Finance*
*Southwestern University of Finance and Economics*
*Chengdu, Sichuan, P. R. China.*
qiangliu@swufe.edu.cn
ORCID: 0000-0001-8466-3108

Yuhan JIAO
*School of Finance*
*Southwestern University of Finance and Economics*
*Chengdu, Sichuan, P. R. China.*
yuhanjiao@smail.swufe.edu.cn
ORCID: 0000-0002-7218-2013

Shuxin GUO[a]
*School of Economics and Management*
*Southwest Jiaotong University*
*Chengdu, Sichuan, P. R. China*
shuxinguo@home.swjtu.edu.cn
ORCID: 0000-0001-8188-6517

a) Correspondence author. School of Economics and Management, Southwest Jiaotong University, No.111, Section 1, North Erhuan Road, Chengdu, Sichuan 610031, P. R. China. Cell: 86-181-0901-5782, Fax: 86-028-8760-0543. Email: shuxinguo@home.swjtu.edu.cn.

**Acknowledgements**

Early on, Q. Liu benefited greatly from in-depth discussions with Timothy F. Crack. This version supersedes the working papers from University Library of Munich entitled "Options Pricing with Arithmetic Brownian Motion and its Implication for Risk-Neutral Valuation" (Dated Dec. 1, 2005) and on SSRN entitled "Options' Prices Under Arithmetic Brownian Motion and Their Implication for Modern Derivatives Pricing" (posted on Jan. 29, 2007). This work was supported by the National Natural Science Foundation of China [Grant number 71701171], and by Huaxi Futures Co., Ltd.

# Risk-Neutral Valuation of Options under Arithmetic Brownian Motions

**ABSTRACT**: On April 22, 2020, the CME Group switched to Bachelier pricing for a group of oil futures options. The Bachelier model, or more generally the arithmetic Brownian motion (ABM), is not so widely used in finance, though. This paper provides the first comprehensive study of options pricing under ABM. Using the risk-neutral valuation, we derive formulas for European options for three underlying types, namely an underlying that does not pay dividends, an underlying that pays a continuous dividend yield, and futures. Further, we derive Black-Scholes-Merton-like partial differential equations, which can in principle be utilized to price American options numerically via finite difference.

**JEL** Classification: G13, G12, A23

# 1 Introduction

Even since the groundbreaking work of Black-Scholes-Merton (BSM) (Black and Scholes, 1973; Merton, 1973), asset price processes have been modeled as geometric Brownian motions (GBM). Much less attention seems to have paid to a closely related, somewhat simpler process, the arithmetic Brownian motion (ABM). There are of course good reasons for this, because ABM allows for negative prices, while it stands to reason that common asset prices cannot be negative.

This widely held belief of non-negativeness was shattered abruptly by the outbreak of the Covid-19 pandemic. "The price of the main U.S. oil benchmark fell more than $50 a barrel to end the day about $30 below zero, the first time oil prices have ever turned negative."[1] The second day, the CME Group stunned the world by issuing an advisory notice with the subject of "Switch to Bachelier Options Pricing Model," mentioning negative strikes for a group of oil futures options contracts for the first time.[2] This unprecedented development brings the Bachelier model back to the attention of both the professionals and academia.

"[In 1900] Louis Bachelier deduced an option pricing formula based on the assumption that stock prices follow a Brownian motion with zero drift" (Merton, 1973). "Unfortunately, Bachelier's assumptions violate some basic economic principles. In particular, he violates limited liability, time preference, and risk aversion" (Crack, 2004). Such a zero-drift Brownian motion is not enough, though. To be more realistic, what we need is a Brownian motion with a non-zero (constant) drift, or the so-called ABM. Even though ABM has not been used to model asset price processes in general, it indeed appears in some other important applications, such as option pricing under ABM with an absorbing barrier at zero to avoid negative underlying prices (Goldenberg, 1991), the forward rate model used in the Heath-Jarrow-Morton model (Heath et al., 1992), and European spread options price formulas under two arithmetic Brownian motions (Poitras, 1998). With the breathtaking CME development, options pricing under ABM imminently becomes a legitimate and important issue worth studying.

---

[1] S. Reed and C. Krauss. "Too Much Oil: How a Barrel Came to Be Worth Less Than Nothing." *New York Times*, April 20, 2020.

[2] CME Group Advisory Notice. CME Clearing Plan to Address the Potential of a Negative Underlying in Certain Energy Options Contracts. Advisory #: 20-152, April 8, 2020.

Within the BSM framework, it is well-known that we can use either the BSM partial differential equation (PDE) approach or the risk-neutral (RN) valuation to obtain price formulas for options. The (second-order) BSM equation is notoriously difficult to solve, while the risk-neutral valuation is much easier to handle. As a result, the RN valuation is usually the first choice whenever possible. Unfortunately, it is not uncommon to make mistakes when applying the RN valuation (see, for example, Kuruc (2003)).

This paper sets out to show both a common wrong and the correct approaches using the RN valuation for an underlying without paying dividends. Then, we generalize the pricing formulas for European options to an underlying with a continuous dividend yield and to futures. Further, we derive BSM-like PDEs, one of which is used to verify both the wrong and correct price formulas. Note that these PDEs can, in principle, be utilized to price American options numerically via finite difference. Finally, we discuss the properties of options prices under ABM as well as two interesting findings of seeming violations of the no-arbitrage principle.

This paper makes three contributions. First, we significantly update the working paper that first utilizing the RN valuation to study the pricing of options under a single underlying variable following ABM, which has been cited by six published scholarly works so far.[3] Second, we are the first to address options pricing by explicitly allowing negative underlying prices. Third, we provide professionals with ready-to-use time-$t$ pricing formulas (as well as PDEs) for three underlying types, and give academics a foundation for further research.

The paper is arranged as follows. In Section 2, we use the risk-neutral valuation to derive both a wrong and the correct price formula for European options with an underlying that does not pay dividends. Section 3 extends the result of Section 2 to an underlying with a continuous dividend yield. The formulas for futures options are presented in Section 4. Section 5 derives Black-Scholes-Merton-like PDEs. Some properties of the pricing formulas are discussed in Section 6. Finally, the paper concludes with further comments.

## 2 Risk-neutral Valuation

The risk-neutral valuation (Cox and Ross, 1976) seems trivial to apply. A three-step procedure for the RN valuation under GBM is outlined in Hull (2015). First, set the expected return of the underlying security to the risk-free interest rate. Second, compute the expected payoff of the derivative at maturity. Third, discount the result of the second step by the risk-free rate to obtain the fair price for an asset.

Unfortunately, this simple RN procedure had been applied incorrectly under arithmetic Brownian motions, which seemed quite common among professional authors (Kuruc, 2003) when we initially set eyes on this issue in 2005. For a better understanding of the problem at hand, we first present a common wrong approach to set the stage for correctly solving the pricing later. Then we describe the correct approach and derive the price formulas for the case of no dividends, the results of which were posted on SSRN in 2007.

---

[3] Liu, Q. 2007. "Options' Prices Under Arithmetic Brownian Motion and Their Implication for Modern Derivatives Pricing." *SSRN*, Jan. 29, 2007. The earliest online version -- Liu, Q. 2005. "Options Pricing with Arithmetic Brownian Motion and its Implication for Risk-Neutral Valuation." *Finance 0512001*, University Library of Munich, Germany -- is available from econwpa.ub.uni-muenchen.de.

**2.1 A Wrong Approach**

According to the procedure outlined above, one can price a derivative by replacing the underlying growth rate in the model of the underlying process by the risk-free interest rate and taking the expectation of the discounted payoff of the derivative. Assume that the underlying price process follows ABM:

$$dS_t = \mu dt + \sigma_S dW_t \tag{1}$$

where $\mu$ is the (constant) drift term, $\sigma_S$ is the (constant) standard deviation of the change in the underlying price (in units of years), and $W_t$ is a Brownian motion under the physical measure. Note that $\sigma_S$ is not the "volatility" known in the finance literature, because (annualized) volatility commonly refers to the standard deviation of the continuously-compounded returns in the underlying. Equation (1) is called a stochastic differential equation (SDE). Further, assume a cash bond model,

$$dP_t = P_t r dt \tag{2}$$

where $P_t$ is the bond price and $r$ is the (constant) risk-free interest rate.

In a typical (or naïve) risk-neutral valuation, $\mu$ is set to $r$ in Equation (1):

$$dS_t = rdt + \sigma_S dB_t \tag{3}$$

where $B_t$ denotes the Brownian motion under the risk-neutral measure. Trivially, the following is obtained by integration:

$$S_T = S_t + r\tau + \sigma_S(B_T - B_t)$$

Note that $B_T - B_t$ is distributed as conditional normal with zero mean and variance of $\tau = T - t$, or $B_T - B_t \sim N(0, \tau)$. Therefore, the price of a European call option is:

$$c_t = e^{-r\tau} E^Q[\max(S_T - K, 0) | S_t]$$
$$= e^{-r\tau} \sigma_S \sqrt{\tau} [d_N N(d_N) + n(d_N)]$$
(4).

$$d_N = \frac{S_t + r\tau - K}{\sigma_S \sqrt{\tau}}$$

where $E$ denotes the expectation operator, $Q$ indicates the risk-neutral measure, $K$ is the strike of the call option, $N(\bullet)$ as usual is the cumulative distribution function (CDF) of the standard normal, and $n(\bullet)$ is the probability density function (PDF) of the standard normal. Similarly, the put price is:

$$p_t = e^{-r\tau} E^Q[max(K - S_T, 0)|S_t]$$
$$= e^{-r\tau} \sigma_S \sqrt{\tau} [n(d_N) - d_N N(-d_N)]$$

**2.2 The Correct Approach**

The formal approach of the risk-neutral valuation also involves three steps. First, transform the discounted underlying process into a martingale under the risk-neutral measure $Q$, using Girsanov's theorem. Second, represent the underlying SDE by this new $Q$-Brownian motion. Third, price any derivative by taking the expectation of the discounted payoff with respect to this $Q$-martingale measure (Baxter and Rennie, 1996; Joshi, 2003).

For the same models as represented by Equations (1-2), form the discounted stock process $Z_t = P_t^{-1} S_t$. The SDE for $Z_t$ is:

$$dZ_t = d(P_t^{-1} S_t) = P_t^{-1}[\sigma_S dW_t + (\mu - rS_t)dt]$$

Let $\sigma_S dB_t = \sigma_S dW_t + (\mu - rS_t)dt$, then $Z_t$ is a martingale under the new measure of $B_t$ (Girsanov's theorem). Putting $B_t$ back into SDE (1), one can obtain the correct risk-neutral SDE as:

$$dS_t = rS_t dt + \sigma_S dB_t \quad (5).$$

Note that the drift coefficient is now $rS_t$, rather than simply $r$ as in Equation (3).[4]

The underlying price at the option's maturity is obtained by solving Equation (5):

$$S_T = e^{r\tau}[S_t + \sigma_S e^{rt} X_T]$$

$$X_T = \int_t^T e^{-rs} dB_s$$

$$X_T \sim N(0, \Sigma^2), \Sigma^2 = \frac{e^{-2rt} - e^{-2rT}}{2r}$$

The Ito integral $X_T$ is distributed as $N(0, \Sigma^2)$, where $\Sigma^2$ is computed using the Ito isometry (Øksendal, 1995; Klebaner, 1998). Clearly, the "simpler" ABM model for the underlying security process leads to more complicated mathematics in pricing derivatives than does the GBM used in the BSM framework. Finally, the call and put prices of European option are, respectively:

$$c_t = \sigma_S \Sigma e^{rt}[dN(d) + n(d)] \qquad (6).$$

$$p_t = \sigma_S \Sigma e^{rt}[n(d) - dN(-d)] \qquad (7).$$

$$d = \frac{S_t - Ke^{-r\tau}}{\sigma_S \Sigma e^{rt}}$$

It is not difficult to verify that Equations (6-7) satisfy the put-call parity (Merton, 1973; Hull, 2015). Note that Formula (6) is given by Crack (2004) via a different approach. In fact, Crack's method, namely solving a second-order BSM-like PDE, is considerably more complicated.[5] If the risk-free rate is zero, the same result as is discussed extensively in Crack (2004) can be obtained. In terms of practical applicability, Formulas (6-7) can be utilized to price spread options.

---

[4] Incidentally, the drift coefficient *mu* in equation (1) divided by the underlying price has the same unit as the growth rate of the underlying. Therefore, it cannot be right to set *mu* directly to *r*. If *mu* is set to $rS_t$ however, the risk-neutral SDE (5) could then be obtained heuristically.

[5] "I derived my solution by solving the ABM PDE subject to the ABM boundary conditions." Crack wrote that that with the help of a mathematician, he spent many hours for the solution (private communication, July 3, 2005).

**3 The Case of Continuous Dividend Yield**

Now we can extend Equations (6-7) to the case when the underlying pays a continuous dividend yield. From Hull (2015), it is known that the RN dynamics is simply:

$$dS_t = (r - q)S_t dt + \sigma_S dB_t \qquad (8).$$

where $q$ denotes the continuous dividend yield. Following the derivation in the Correct Approach section, we have the terminal underlying price as:

$$S_T = e^{(r-q)\tau}[S_t + \sigma_S e^{(r-q)t} X_T]$$

$$X_T = \int_t^T e^{-(r-q)s} dB_s$$

$$X_T \sim N(0, \Sigma_q^2), \Sigma_q^2 = \frac{e^{-2(r-q)t} - e^{-2(r-q)T}}{2(r-q)}$$

The call and put option prices are:

$$c_t = e^{-q\tau} h[d_q N(d_q) + n(d_q)] \qquad (9).$$

$$p_t = e^{-q\tau} h[n(d_q) - d_q N(-d_q)] \qquad (10).$$

$$h = \sigma_S \Sigma_q e^{(r-q)t}, d_q = \frac{S_t - Ke^{-(r-q)\tau}}{h}$$

Poitras (1998) derives a complex formula for European spread options with two arithmetic Brownian motions, which reduces to Formula (9) with $t = 0$ under rather tight conditions. Again, it is not difficult to show that Equations (9-10) satisfy the put-call parity (Merton, 1973; Hull, 2015). Further, it is easy to verify that when $q$ is zero, Equation (9) reduces to (6), and Equation (10) to (7). It is worth pointing out that Equations (9-10) are also applicable to foreign exchange options, if $q$ is replaced by the foreign risk-free interest rate, and to futures options (see the next section).

**4 Futures Options**

CME switched to Bachelier pricing for a list of oil futures options. Therefore, it is certainly necessary to provide pricing formulas for futures options. With our previous general results, we can easily derive those formulas now. In fact, there are at least two ways to do this.

First, futures prices follow the simple RN process, $dF_t = \sigma_F dB_t$ (Hull, 2015). This is the basis of the famous Bachelier model (Merton, 1973; Goldenberg, 1991). One can then easily obtain the options price formulas by solving this SDE and applying the RN valuation. Note that this SDE is the same as Equation (3) with zero $r$. Therefore, we have from Formula (4) with zero $r$ in $d_N$:

$$c_t = e^{-r\tau} \sigma_F \sqrt{\tau} [d_F N(d_F) + n(d_F)]$$

$$d_F = \frac{F_t - K}{\sigma_F \sqrt{\tau}}$$

The formula for put is omitted here, because it can be trivially obtained via the put formula in exactly the same way.

Second, it is known that futures can be regarded as an underlying that pays a continuous dividend yield (Hull, 2015). Therefore, the same result for calls can be obtained from Formula (9) by equating $q$ to $r$. Note that the limit of $\Sigma_q^2$ in Formula (9) is $\tau$, when $q$ approaches $r$. Of course, the pricing formula for puts can be obtained via Formula (10) in exactly the same way.

**5 Partial Differential Equations**

Intuitively, we know that the call price in (4) is incorrect, while Equation (6) is correct. An interesting question now is: how can we verify this? To answer this question, we can, fortunately, utilize the Black-Scholes-Merton PDE approach, because it is well-known that the risk-neutral valuation and the PDE method are equivalent within the modern options pricing theory of BSM (Hull, 2015; Neftci, 2000). Formally, following the procedure of BSM, we can derive BSM-like PDEs and then solve them to obtain the prices of options.

For easy exposition, we start with the general case of continuous dividend yield. Given the underlying model in Equation (8), a BSM-like PDE can be obtained by forming a suitable portfolio of the option and its underlying, applying Ito's lemma, and eliminating the random Brownian movement. For notational brevity, we drop the subscript for time here:

$$\frac{\partial v}{\partial t} + (r-q)S\frac{\partial v}{\partial S} + \frac{1}{2}\sigma_S^2 \frac{\partial^2 v}{\partial S^2} = rv \qquad (11).$$

where *v* denotes the price of an option (see the Appendix for the derivation). Compared with the corresponding BSM equation, the only difference is the missing of the $S^2$ factor in front of the second-order partial derivative. Importantly, the drift coefficient for the underlying process does not appear in the PDE, just as in the original BSM equation. Therefore, the risk-neutral valuation is also equivalent to the PDE approach, and indeed applicable in the case of ABM.

It is quite a challenge to solve equation (11) directly to obtain a fair price for a European option. Fortunately, this does not have to be done here, however. Remember that we have already obtained formulas for option prices, which can be simply plugged into Equation (11) to see whether any is a solution of it. The correct price should satisfy Equation (11), just as in the case of BSM.

Formulas (4) and (6) will be checked as our examples here. By setting *q* to be zero in (11), we obtain the following PDE:

$$\frac{\partial v}{\partial t} + rS\frac{\partial v}{\partial S} + \frac{1}{2}\sigma_S^2 \frac{\partial^2 v}{\partial S^2} = rv$$

It turns out that, after some length algebraic manipulations, which are omitted from this paper, Formula (6) is found to solve the PDE, while (4) does not.[6] Therefore, Formula (4) from the wrong approach is indeed incorrect. This result is reassuring and expected, since the formal approach should intuitively provide us with the correct answer. As a matter of fact, the equivalent of the conditional expectations used to obtain Formulas (9-10) and the PDE (11) can be easily verified from the well-known Kolomogorov's Equation or Feynman-Kac Formula (Klebaner, 1998; Neftci, 2000; Øksendal, 1995). Note that when the risk-free rate is zero, the prices (4) and (6) do become the same, which is expected because in this case the SDEs (3) and (5) are the same to begin with.

Note that Formula (4) is incorrect for a second reason, because it and the corresponding put formula do not satisfy the put-call parity. On the other hand, Formulas (6-7), as well as (9-10), are consistent with the put-call parity.

Finally, because $q$ for futures is equal to $r$, we have the following simple PDE for futures options:

$$\frac{\partial v}{\partial t} + \frac{1}{2}\sigma_S^2 \frac{\partial^2 v}{\partial S^2} = rv$$

which once again is very similar to the corresponding PDE under GBM.

These BSM-like PDEs cannot be solved analytically for American options, just as the BSM PDEs cannot. Fortunately, they can be solved numerically to obtain prices for American options.

**6 Properties of Option Prices Under ABM**

**6.1 Call Price**

It is well-known that the Black-Scholes-Merton formulas have correct properties in all limiting cases. It is interesting to ask now whether the pricing formulas under ABM have the same properties.

---

[6] Timothy F. Crack pointed out that it is necessary, but not sufficient, for the price (6) to satisfy the PDE. Crack asserted that this solution is correct, however (private communication, July 3, 2005).

As an example, we choose the simpler version, Formula (6) for call prices, to work with. Put prices can be analyzed similarly, and will not be described further here. When the underlying price is very high, *d* becomes very big as well. As a result, *N(d)* approaches one while *n(d)* becomes negligible, and thus,

$$c_t \to S_t - Ke^{-r\tau}$$

which is the price of a stock forward and the correct limit. On the other hand, if the stock price approaches zero, *d* becomes a negative number on the order of the strike (given typical values of risk-free rate, $\sigma_S$, and maturity). Thus, both *N(d)* and *n(d)* are close to zero, so the call price is close to zero. This is again the expected behavior (Hull, 2015).

For very small $\sigma_S$, if $S_t - Ke^{-r\tau}$ is positive, then *d* becomes very large, and again $c_t \to S_t - Ke^{-r\tau}$; if $S_t - Ke^{-r\tau}$ is negative, *d* approaches negative infinity, and both *N(d)* and *n(d)* are close to zero, so the call price is close to zero. These are correct, according to Hull (2015) once again.

Finally, when it is very close to maturity, $\Sigma$ approaches zero. For positive $S_t - K$, *d* becomes very big and the call price is simply $S_t - K$; otherwise, the call price is zero. These are, by definition, the two parts of the payoff of European call options.

In summary, the call price formula has the correct limiting properties in all the extreme cases above. Further, all these results can be confirmed by numerical calculations.

**6.2 Numerical Examples**

Let us show a few numerical results for easy understanding. From SDE (1), it is easy to see that the estimated $\hat{\sigma}_S$ is $(Var[dS_t]/dt)^{0.5}$, where *Var* denotes variance. Note that the variance here is computed using daily price differences, instead of the daily logarithm returns, and thus depends on the level of prices. Using the S&P 500 index in 2019 as an example, we obtain $\hat{\sigma}_S = 354.77$. For comparison, the annualized volatility as we commonly know is only 12.5%. Consequently, a direct numerical comparison between options pricing under ABM and GBM is difficult, if not impossible.

Obviously, we can see $\sigma_S$ in SDE (1) is roughly the product of volatility and stock price in GBM. For an average price 10 and a volatility 30%, one has $\sigma_S = 3$. Further, assume a risk-free rate of 5%, and a maturity of 0.5 years. Finally, $t = 0$ for today and $T = \tau$ in years. With these parameters, the prices from Formulas (6-7) are shown in Figure 1 with a strike of 5 and Figure 2 with a strike of -5, respectively.

Figure 1 here

Figure 2 here

Several observations can be made here. Both the call and put prices are convex functions of the underlying price, just as the BSM formulas are. The call price with a positive strike is small (but not zero), when the underlying price is zero (Figure 1). With a negative strike, the call price can be quite large at 4.88, when the underlying price is zero (Figure 2). The price curves in Figures (1-2) look very similar, appearing roughly shifted horizontally with the strike.

When maturity goes to infinity (i.e., a perpetual call or warrant), the result is more complicated and shown in the following:

$$c_t \to S_t N(\frac{\sqrt{2r}}{\sigma_S} S_t) + \frac{\sigma_S}{\sqrt{2r}} n(\frac{\sqrt{2r}}{\sigma_S} S_t)$$

For typical $\sigma_S$ and high stock prices, $N(\bullet)$ is close to one while $n(\bullet)$ is to zero, and thus the call price approaches the stock price. This is the expected behavior due to no arbitrage (Merton, 1973). For very low stock prices, however, the call price approaches $\sigma_S/(2\sqrt{r\pi})$, a finite positive number, which can be seen clearly even with a five-year maturity in Figure 3.

Figure 3 here

It is well-known that for an underlying with limited liability, the underlying price is the upper bound of the call price (Merton, 1973). To put it simply, if a call price is above the stock price, one can sell the call, use some of the proceeds to buy the stock (hence a covered call), and invest the rest of the proceeds in a money market account. At maturity, the covered call will be worth either the strike (in-the-money) or the stock (out-of-the-money). Either way, the strategy will make money without any initial costs, which is termed a riskless arbitrage. In Figure 3, however, the call price crosses the upper bound at a stock price above 1. This seems violating the no-arbitrage upper bound of the call price, in case the underlying is of limited liability.[7]

In addition, if $\sigma_S$ approaches infinity, $d$ approaches zero when the stock price is low. The call price tends to a finite $\sigma_S \Sigma e^{rt}/\sqrt{2\pi}$. This again appears violating the no-arbitrage upper bound of a European call. Even for a moderately high $\sigma_S$ of 12, the call price crosses the upper bound at roughly a stock price of 2 (see Figure 4).

Figure 4 here

Now one may wonder why the call price derived from the risk-neutral valuation under ABM seems to violate the principle of no arbitrage. It is well-known that the BSM formulas under GBM do not violate the no-arbitrage condition in any extreme case. One might argue that ABM is not a good model for stock price dynamics because stock prices cannot be negative. This point is debatable at best, however. Bouchaud and Potters (2000) demonstrate that the approximation of ABM is almost as good or bad as that of GBM. Furthermore, the observed volatilities tend to be higher at lower stock prices (Demeterfi et al., 1999). Actually, a constant $\sigma_S$ under ABM implies that the volatility for the same stock is higher for lower prices, while volatility under GBM is constant regardless of stock price levels. Therefore, ABM may arguably be a better model.

---

[7] We thank Joseph Pimbley for helping make the statement correct.

The two cases of seeming violations of no arbitrage under ABM is no doubt an intriguing phenomenon. We argue that if the underlying price is allowed to be negative, it is no longer the upper bound of the call price. Because a call option is always of limited liability, its value can never be negative. For a negative underlying price, then the call prices are certainly higher. Therefore, higher call prices in Figures (3-4) over the underlying prices, as the results of unlimited liability of the underlying, are correct and in fact do not violate the principle of no arbitrage.

## 7 Conclusions

Using the risk-neutral valuation, this paper derives option pricing formulas under arithmetic Brownian motions (ABM). We provide call and put formulas for three underlying types, namely the underlying without dividend, the underlying with a continuous dividend yield, and futures.

Further, this paper derives Black-Scholes-Merton-like partial differential equations (PDE) under ABM. We use one of the PDEs to verify the pricing formulas. Those PDEs can be solved numerically via finite difference to price American options.

Finally, this paper discusses properties of option pricing formulas under ABM. Encouragingly, those formulas allow both negative underlying and strike prices, as per the CME requirements. Surprisingly though, these formulas appear to violate the principle of no arbitrage in at least two situations, namely long maturities and high standard deviation of the change in underlying prices. Under ABM or unlimited liability, those violations do not amount to true arbitrage, however.

**Appendix**

Assume that the price of options is a function of the underlying price and time, $v(S, t)$. Applying Ito's lemma, we have:

$$dv = (\frac{\partial v}{\partial t} + \mu \frac{\partial v}{\partial S} + \frac{1}{2}\sigma_S^2 \frac{\partial^2 v}{\partial S^2})dt + \sigma_S \frac{\partial v}{\partial S} dW$$

For a portfolio consisting one long call option and $\Delta$ short underlying at time $t$, the value is:

$$\pi = v - \Delta S$$

With the passing of infinitesimal time of $dt$, the value of the portfolio changes instantaneously by an amount of,

$$d\pi = dv - \Delta dS$$

$$= (\frac{\partial v}{\partial t} + \mu[\frac{\partial v}{\partial S} - \Delta] + \frac{1}{2}\sigma_S^2 \frac{\partial^2 v}{\partial S^2})dt + \sigma_S[\frac{\partial v}{\partial S} - \Delta]dW$$

which will be riskless if $\Delta = \partial v / \partial S$.

In addition to the value change of the portfolio, the investor has to pay the stock owner the dividend yield continuously over time $dt$. Thus, the rate of net wealth change will be certain (or riskless) given $\Delta = \partial v / \partial S$, and is equal to the risk-free interest rate on $\pi$:

$$(\frac{\partial v}{\partial t} + \frac{1}{2}\sigma_S^2 \frac{\partial^2 v}{\partial S^2})dt - \Delta q S dt = r\pi dt$$

Plugging in $\Delta = \partial v / \partial S$, we have:

$$\frac{\partial v}{\partial t} - \frac{\partial v}{\partial S} qS + \frac{1}{2}\sigma_S^2 \frac{\partial^2 v}{\partial S^2} = rv - \frac{\partial v}{\partial S} rS$$

Rearranging the terms, we obtain Equation (11).

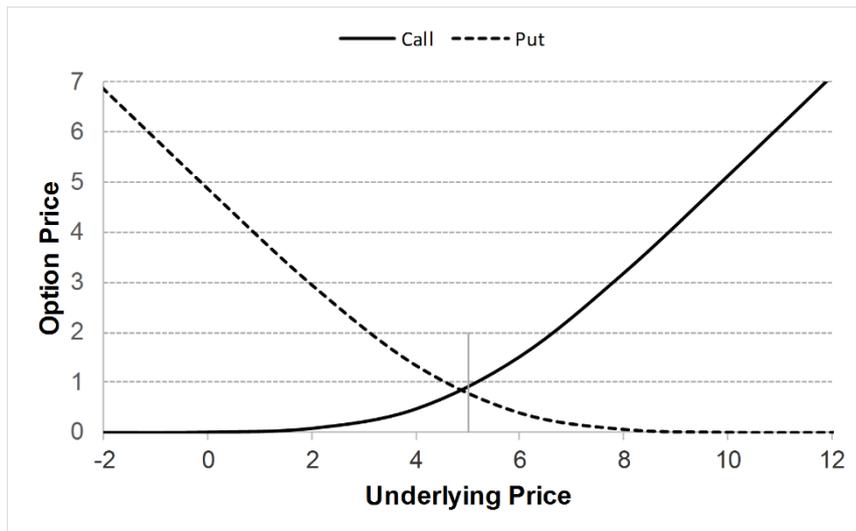

**Figure 1**. **The Price of Options as a Function of the Underlying Price, Which Follows an Arithmetic Brownian Motion.** The strike price is 5, and indicated by a vertical line segment in the middle. $\sigma_S = 3$. The risk-free rate is 5%. The maturity is 0.5 years.

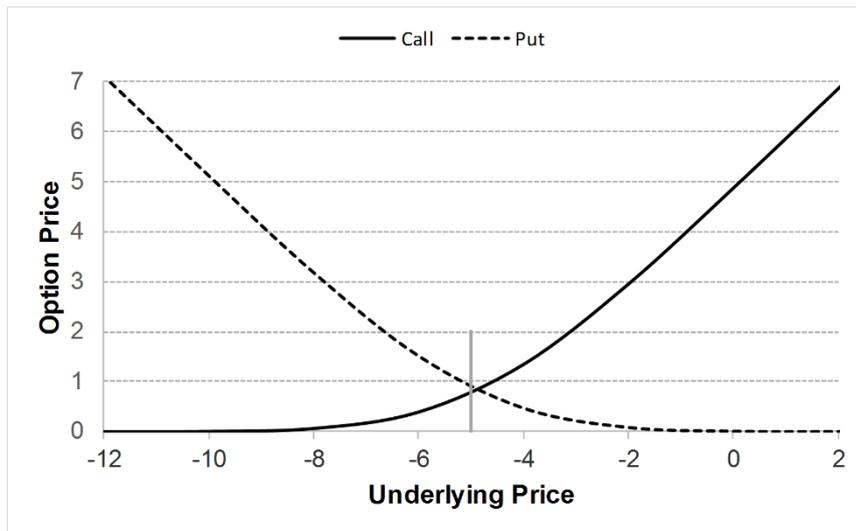

**Figure 2**. **The Price of Options as a Function of the Underlying Price, Which Follows an Arithmetic Brownian Motion.** The strike price is negative 5, and indicated by a vertical line segment in the middle. $\sigma_S = 3$. The risk-free rate is 5%. The maturity is 0.5 years.

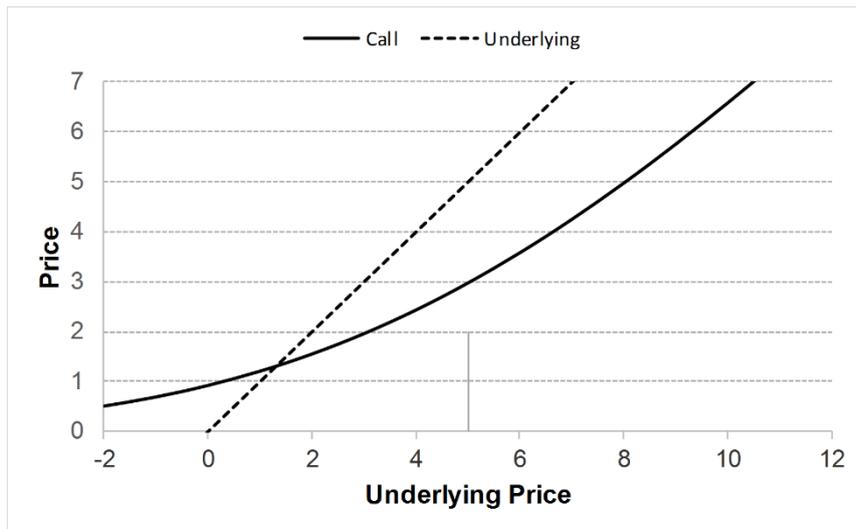

**Figure 3**. **The Price of an Underlying or a Call Option as a Function of the Underlying Price, Which Follows an Arithmetic Brownian Motion.** The strike price is 5, and indicated by a vertical line segment in the middle. $\sigma_S = 3$. The risk-free rate is 5%. The maturity is 5 years.

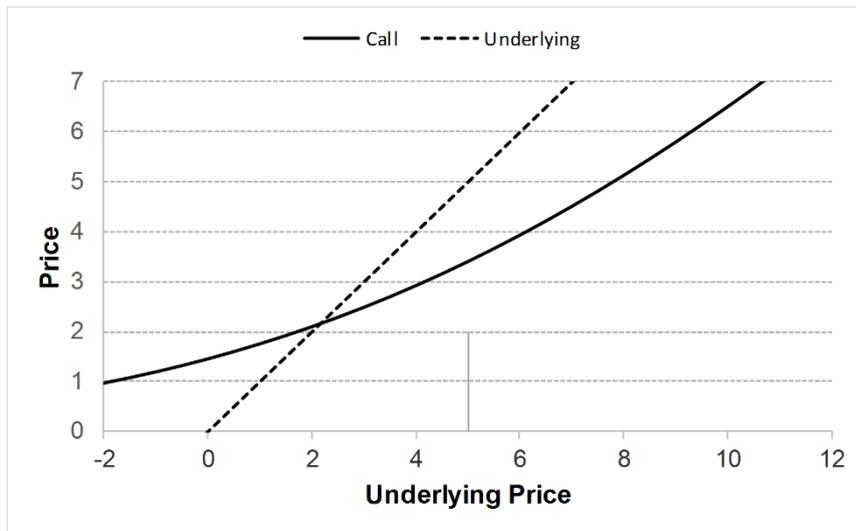

**Figure 4**. **The Price of an Underlying or a Call Option as a Function of the Underlying Price, Which Follows an Arithmetic Brownian Motion.** The strike price is 5, and indicated by a vertical line segment in the middle. $\sigma_S = 12$. The risk-free rate is 5%. The maturity is 0.5 years.